\documentclass[fleqn,10pt]{wlscirep}
\usepackage[utf8]{inputenc}
\usepackage[T1]{fontenc}
\usepackage{multirow}
\usepackage{soul}
\usepackage{cancel}

\title{Evolution of direct reciprocity in group-structured populations}

\author[1,*]{Yohsuke Murase}
\author[2]{Christian Hilbe}
\author[3]{Seung Ki Baek}
\affil[1]{RIKEN Center for Computational Science, Kobe, Hyogo 650-0047, Japan}
\affil[2]{Max Planck Research Group `Dynamics of Social Behavior', Max Planck Institute for Evolutionary Biology, Pl\"on, Germany}
\affil[3]{Department of Scientific Computing, Pukyong National University, Busan 48513, Korea}

\affil[*]{yohsuke.murase@gmail.com}

\begin{abstract}
People tend to have their social interactions with members of their own community.
Such group-structured interactions can have a profound impact on the behaviors that evolve. 
Group structure affects the way people cooperate, and how they reciprocate each other's cooperative actions. 
Past work has shown that population structure and reciprocity can both promote the evolution of cooperation. 
Yet the impact of these mechanisms has been typically studied in isolation. 
In this work, we study how the two mechanisms interact. 
Using a game-theoretic model, we explore how people engage in reciprocal cooperation in group-structured populations, compared to well-mixed populations of equal size. 
To derive analytical results, we focus on two scenarios. 
In the first scenario, we assume a complete separation of time scales. 
Mutations are rare compared to between-group comparisons, which themselves are rare compared to within-group comparisons. 
In the second scenario, there is a partial separation of time scales, where mutations and between-group comparisons occur at a comparable rate. 
In both scenarios, we find that the effect of population structure depends on the benefit of cooperation. 
When this benefit is small, group-structured populations are more cooperative. 
But when the benefit is large, well-mixed populations result in more cooperation. 
Overall, our results reveal how group structure can sometimes enhance and sometimes suppress the evolution of cooperation. 
\end{abstract}
\begin{document}

\flushbottom
\maketitle
\thispagestyle{empty}


\section*{Introduction}

\begin{figure}
    \centering
    \includegraphics[width=0.6\textwidth]{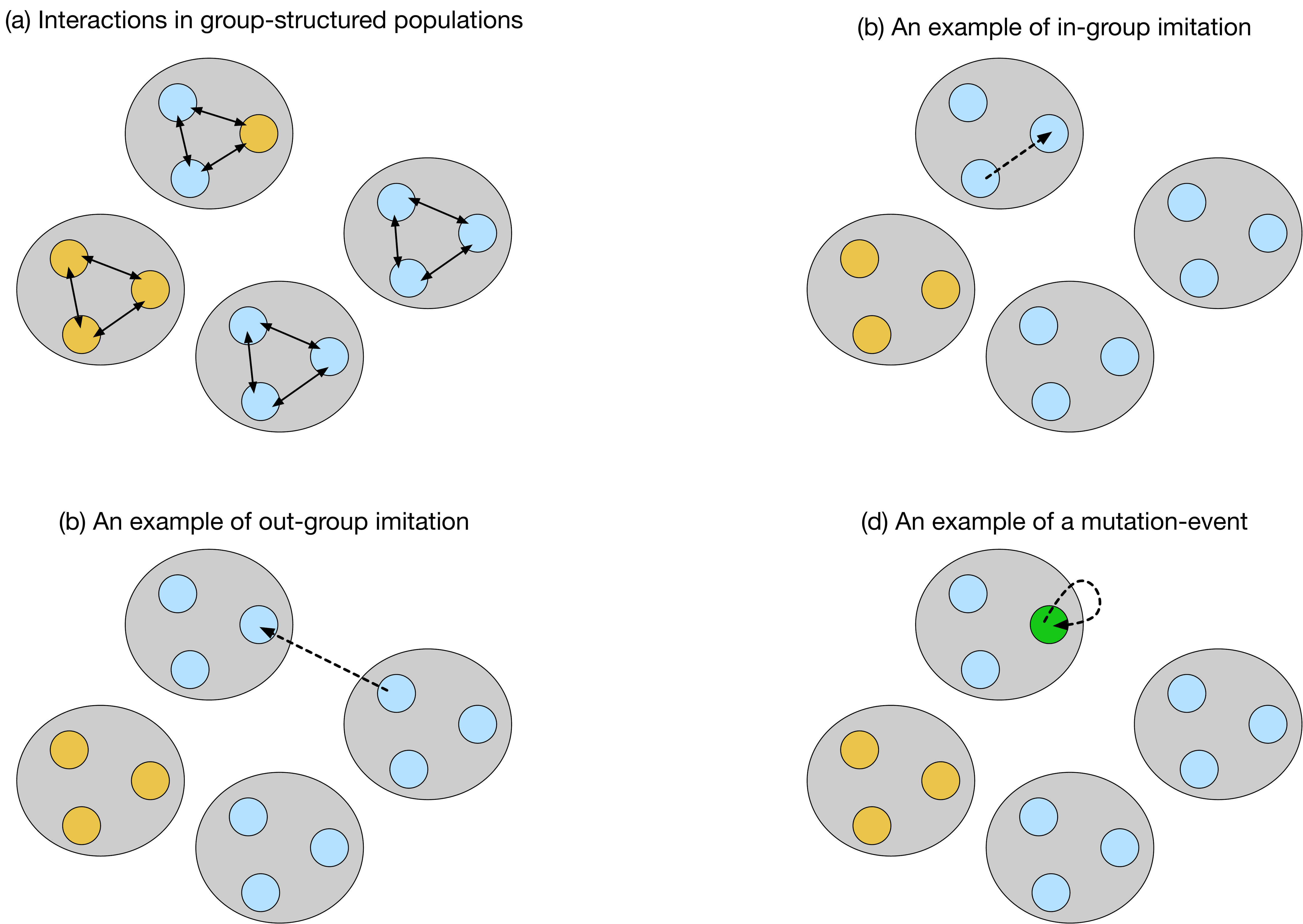}
    \caption{{\bf A schematic representation of the model setting.}
    (a) We consider pairwise interactions in group-structured populations. The population is composed of $M$ groups (depicted by grey sets). Each group contains $N$ players (depicted by colored circles). In this example, $M=4$ and $N=3$, such that the total population size is $NM=12$.
    The colors of the small circles indicate the players' strategies.
    The black arrows between these circles indicate that each player interacts in a repeated prisoner's dilemma with all other group members.
    Over time, players may change their strategies for the repeated prisoner's dilemma in three different ways. 
    (b) First, with probability $\mu_{\rm in}$, players engage in intra-group imitation. In that case, a player randomly samples a role model from the same group, and imitates the role model's strategy with a certain probability that depends on the players' payoffs. 
    (c) Second, with probability $\mu_{\rm out}$, players engage in inter-group (or out-group) imitation. In that case, the player randomly selects a role model from a different group, and again imitates this role model with a certain probability. 
    (d) Finally, with probability~$\nu$, there is a mutation, in which case the player selects to a new strategy randomly. The three probabilities sum up to one, $\mu_{\rm in} + \mu_{\rm out} + \nu = 1$. When we consider a scenario with a complete separation of time scales, we assume $\nu \ll \mu_{\rm out} \ll \mu_{\rm in}$. When we consider a partial separation of time scales, we assume $\nu\ll \mu_{\rm in}$ and $\mu_{\rm out} \ll \mu_{\rm in}$.}
    \label{fig:model}
\end{figure}

Human populations have some internal structure~\cite{nowak2010evolutionary,albert2002statistical}. Even in our modern and highly connected societies, the number of meaningful social ties that an individual can have is limited~\cite{dunbar1992neocortex,dunbar1993coevolution}. These limitations in turn restrict the social interactions that are possible. Yet most models of reciprocal cooperation do not consider these restrictions; they assume populations are well-mixed\cite{axelrod:Science:1981,sigmund:book:2010,hilbe2018partners}. All individuals of a population are equally likely to interact with each other, and equally likely to imitate each other's strategies. In the following, we explore how these results on reciprocity in well-mixed populations generalize to populations with a group structure. Following the work of Hauert, Chen, and Imhof on one-shot games\cite{hauert2012evolutionary,hauert2014fixation}, we consider a situation in which a population is subdivided into smaller groups. Each individual in such a group engages in a repeated game with every other group member, as displayed in Fig.~\ref{fig:model}{\color{blue}a}. To play these games, individuals can choose between different memory-1 strategies of direct reciprocity, such as AllD, Tit-for-Tat, or Win-Stay Lose-Shift\cite{kraines1989pavlov,nowak1993strategy}. As time passes by, individuals are not restricted to stick to their respective strategies. Instead, they may adapt their behaviors, by imitating the strategies of other population members with higher payoffs. We assume that these imitation events are most likely to take place within an individual's own group (Fig.~\ref{fig:model}{\color{blue}b}). In addition, there is also some chance that individuals imitate the behaviors of out-group members (Fig.~\ref{fig:model}{\color{blue}c}). In this way, successful behaviors can spread from one group to another, giving rise to a dynamics that is reminiscent of multilevel selection models~\cite{traulsen2006evolution}.

While both population structure and direct reciprocity can promote cooperative behavior on their own\cite{nowak2006five}, it is less obvious what their joint effect is. To see this point, let us recall that many well-known strategies of direct reciprocity can be categorized into two classes, ``partners'' and ``rivals''~\cite{hilbe2018partners,hilbe2015partners}.
Partners denote a set of generous strategies that aim to achieve full cooperation with any given co-player. If the benefit of cooperation is sufficiently large, this class includes, for example, the well-known Win-Stay Lose-Shift rule\cite{kraines1989pavlov,nowak1993strategy}. 
In contrast, individuals with a rival strategy aim to outperform their co-player. Such individuals want to ensure that their own payoff never falls below the co-player's. This class includes extortionate strategies\cite{press2012iterated,mcavoy2016autocratic,szolnoki2014defection,Xu:PRE:2017,ichinose2018zero,Ueda:RSOS:2021} as well as unconditional defectors. 
A series of theoretical studies have revealed whether partners or rivals are favored by selection~\cite{hilbe2014cooperation,stewart2013extortion,stewart2014collapse,stewart2016small,stewart2016evolutionary,akin2015you,akin2016iterated,adami2013evolutionary,hilbe2013evolution,hilbe2013adaptive,baek2016comparing}.
When populations are large and the benefit of cooperation is high, partner strategies are favored. The resulting populations are highly cooperative. 
By contrast, when either the population size or the benefit of cooperation is small, rival strategies are selected, which in turn prevent the evolution of cooperation.

Taking these characteristics of partner and rival strategies into account, the effect of group structure on cooperation is ambivalent.
On the one hand, group structure generally favors cooperation, since individuals from cooperative groups are more likely to be imitated by other population members. 
On the other hand, group structure may reduce the chance that such cooperative groups emerge in the first place. 
By splitting a well-mixed population into several groups, each player experiences a smaller effective population size. Because small population sizes favor the evolution of rival strategies, the overall effect of group structure on cooperation may be negative. 

To study the effect of group structure, we explore the evolution of direct reciprocity in two scenarios:
First, we study the evolutionary dynamics of group-structured populations when there is a complete separation of time scales. 
Here, individuals are most likely to adopt new strategies by imitating another group member (Fig.~\ref{fig:model}{\color{blue}b}), far less likely by imitating an out-group member (Fig.~\ref{fig:model}{\color{blue}c}), and yet again far less likely by exploring a new strategy at random (akin to a mutation in biological models, Fig.~\ref{fig:model}{\color{blue}d}). As a result of this assumption, the population is typically homogeneous, such that all players apply the same strategy. Only occasionally, a new strategy arises by mutation. This new strategy either goes extinct or takes over the population by the time the next mutation arises. 
As our second scenario, we explore a model with a partial separation of time scales. Here, individuals are still most likely to adopt new strategies by imitating another group member. However, in this scenario, out-group imitation and mutations occur at a comparable rate. This assumption implies that  most of the time, each group is still homogeneous, but players from different groups may now use different strategies. Our results suggest that in both scenarios, the effect of group structure is indeed non-trivial. Group structure promotes cooperation when the benefit of cooperation is small. Yet for large benefits, cooperation evolves more easily in well-mixed populations. 

\section*{Model}

The following description of our model consists of two parts. First, we describe how individuals engage in pairwise interactions within their groups. These interactions take the form of an infinitely repeated prisoner's dilemma. We derive the payoffs that each player obtains, given the players' strategies. In a second step, we describe how individuals update their strategies over time, based on a pairwise comparison process\cite{traulsen2006stochastic}.\\

\noindent
{\bf Dynamics of the repeated prisoner's dilemma.} The repeated Prisoner's Dilemma (PD) is the most fundamental theoretical framework to study direct reciprocity. 
The game takes place among two players. 
In each of infinitely many rounds, each player independently decides whether to cooperate ($C$) or to defect ($D$).
In this paper, we study a special variant of the prisoner's dilemma, the donation game. 
Here, cooperation means that a player pays a cost (normalized to $c=1$) in order for the other player to receive a benefit $b\!>\!1$. The respective payoff matrix is
\begin{equation}\label{eq:donation_game}
    \begin{pmatrix}
    b-1 & -1 \\
    b & 0
    \end{pmatrix}.
\end{equation}
If the donation game is only played once, the only Nash equilibrium is to mutually defect. 
But in the repeated game it can be reasonable to cooperate, because players may get a higher long run payoff by maintaining a good relationship with their co-player.
In particular, the Folk theorem of repeated games guarantees that mutual cooperation can be sustained in a Nash equilibrium, provided there are sufficiently many rounds~\cite{fudenberg1991game}.

In general, strategies for the repeated prisoner's dilemma need to tell the player what to do after any history of previous interactions. The resulting space of  
possible strategies is vast\cite{hilbe2017memory}. To simplify our analysis, in the following we focus on a well-known subspace, the space of memory-1 strategies\cite{baek2016comparing}. When players adopt a memory-1 strategy, they only condition their next decision on what happened in the previous round. Such strategies can be represented as a 4-tuple, 
\begin{equation}
\mathbf{p} = (p_{CC}, p_{CD}, p_{DC}, p_{DD}). 
\end{equation}
An entry $p_{ij}$ represents the player's cooperation probability, given that in the previous round the player used action~$i$ and the co-player used action~$j$. We call a memory-1 strategy deterministic, if all entries are either zero or one, and we refer to the set of all such strategies as $\mathcal{M}$. It follows that there are $|\mathcal{M}| = 2^4 = 16$ such strategies in total. They are summarized in Table~\ref{tab:memory-1}. In particular, this table includes several well-known strategies such as AllD, Tit-for-Tat (TFT), or Win-Stay-Lose-Shift~(WSLS)~\cite{kraines1989pavlov,nowak1993strategy}. For our subsequent analysis, we shall assume that players make errors with some small probability $e$. That is, with probability $e$ a player defects although this player intended to cooperate (and conversely, a player who intended to defect may cooperate). As a result, instead of their intended strategies $\mathbf{p}$, players implement the effective strategies $(1\!-\!\varepsilon)\mathbf{p}+\varepsilon(\mathbf{1}-\mathbf{p})$.

When both players adopt a memory-1 strategy, one can explicitly compute their payoffs and how often they cooperate, by representing the game as a Markov chain\cite{sigmund:book:2010}. The possible states of this Markov chain are the possible outcomes of each round (CC, CD, DC, DD; here, the first and second letter refer to the action of the first and second player, respectively). Given these states and the players' (effective) strategies $\mathbf{p} = (p_{CC}, p_{CD}, p_{DC}, p_{DD})$ and $\mathbf{q} = (q_{CC}, q_{CD}, q_{DC}, q_{DD})$, the transition matrix $T$ of the Markov chain takes the form
\begin{equation}
T = \left(
\begin{array}{llll}
p_{CC} \cdot q_{CC} 	&p_{CC} \cdot \bar{q}_{CC}	&\bar{p}_{CC} \cdot q_{CC}	&\bar{p}_{CC}\cdot \bar{q}_{CC}\\
p_{CD} \cdot q_{DC} 	&p_{CD} \cdot \bar{q}_{DC}	&\bar{p}_{CD}\cdot q_{DC}	&\bar{p}_{CD}\cdot \bar{q}_{DC}\\
p_{DC} \cdot q_{CD} 	&p_{DC} \cdot \bar{q}_{CD}	&\bar{p}_{DC}\cdot q_{CD}	&\bar{p}_{DC}\cdot \bar{q}_{CD}\\
p_{DD} \cdot q_{DD} 	&p_{DD} \cdot \bar{q}_{DD}	&\bar{p}_{DD}\cdot q_{DD}	&\bar{p}_{DD}\cdot \bar{q}_{DD}\\
\end{array}
\right).
\end{equation}
Here, $\bar{p}_{ij} = 1-p_{ij}$ and $\bar{q}_{ij} = 1-q_{ij}$ for $i,j\!\in\!\{C,D\}$ are the players' respective defection probabilities. For positive error rates, every entry of this transition matrix is positive. It follows by the Theorem of Perron-Frobenius that $T$  has a unique invariant distribution $\mathbf{v} = (v_{CC}, v_{CD}, v_{DC}, v_{DD})$. 
In particular, the $\mathbf{p}$-player's average cooperation level is $\gamma_{\mathbf{p},\mathbf{q}} := v_{CC}\!+\!v_{CD}$ whereas the $\mathbf{q}$-player's cooperation level is $\gamma_{\mathbf{q},\mathbf{p}}:=v_{CC}\!+\!v_{DC}$. 
As a result, the $\mathbf{p}$-player's long-term average payoff is given by
\begin{equation}\label{eq:long_term_payoff}
   \pi_{\mathbf{p},\mathbf{q}}=  b\cdot \gamma_{\mathbf{q},\mathbf{p}} - c\cdot \gamma_{\mathbf{p},\mathbf{q}}. 
\end{equation}
When both players adopt the same strategy $\mathbf{p}\!=\!\mathbf{q}$ and errors are rare, $e\rightarrow 0$, Table~\ref{tab:memory-1} shows the  cooperation levels $\gamma_{\mathbf{p},\mathbf{p}}$ for each deterministic memory-1 strategy. In particular, the table reveals that there are only three deterministic strategies that fully cooperate against themselves in the limit of rare errors. These strategies are AllC, WSLS, and the strategy $\mathbf{p}\!=\!(1,1,1,0)$. Out of these three strategies, WSLS is a Nash equilibrium if $b\!\ge\!2c$, whereas the other two strategies are always unstable\cite{sigmund:book:2010}.\\ 

\begin{table}[t]
    \centering
    \begin{tabular}{|l|c|c|c|c|}
        \hline
        ID & prescriptions ($CC,CD,DC,DD$) & cooperation level & rival & vs WSLS  \\ \hline
        $S_0$ (AllC)  & $C,C,C,C$                     & 1         &            & $-$ \\
        $S_1$         & $D,C,C,C$                     & 1/2       &            & $-$ \\
        $S_2$         & $C,D,C,C$                     & 3/4       &            & $0$ \\
        $S_3$         & $D,D,C,C$                     & 1/2       &            & $0$ \\
        $S_4$         & $C,C,D,C$                     & 3/4       &            & $-$ \\
        $S_5$         & $D,C,D,C$                     & 1/2       &            & $0$ \\
        $S_6$ (WSLS)  & $C,D,D,C$                     & 1         &            & $0$ \\
        $S_7$         & $D,D,D,C$                     & 1/2       &            & $+$ \\
        $S_8$         & $C,C,C,D$                     & 1         &            & $-$ \\
        $S_9$         & $D,C,C,D$                     & 0         &            & $-$ \\
        $S_{10}$ (TFT)  & $C,D,C,D$                     & 1/2       & \checkmark & $0$ \\
        $S_{11}$       & $D,D,C,D$                     & 1/4       & \checkmark & $0$ \\
        $S_{12}$        & $C,C,D,D$                     & 1/2       &            & $0$ \\
        $S_{13}$        & $D,C,D,D$                     & 1/4       &            & $+$ \\
        $S_{14}$ (GRIM) & $C,D,D,D$                     & 0         & \checkmark & $+$ \\
        $S_{15}$ (AllD) & $D,D,D,D$                     & 0         & \checkmark & $+$ \\
        \hline
    \end{tabular}
    \caption{{\bf Deterministic memory-1 strategies of the repeated prisoner's dilemma.} Memory-1 strategies only depend on the outcome of the previous round. These strategies are deterministic if, given the previous outcome, they either prescribe to cooperate with certainty~($C$), or to defect with certainty~($D$). The four letters in the second column represent which action the strategy prescribes, given the previous actions of the focal player (first letter) and the co-player (second letter).
    The third column shows the cooperation level when both players use the respective strategy (assuming $e \to 0$).
    The fourth column indicates whether or not the strategy is a rival strategy. A strategy $\mathbf{p}$ is a rival if it enforces the payoff relationship $\pi_{\mathbf{p},\mathbf{q}}\ge \pi_{\mathbf{q},\mathbf{p}}$ against all co-player's strategies $\mathbf{q}$. 
    The fifth column shows how each strategy $\mathbf{p}$ performs against WSLS (again for $e\to 0$).
    If $\pi(\mathbf{p},WSLS) > \pi(WSLS,\mathbf{p})$, then we mark the strategy $\mathbf{p}$ with a `$+$'. 
    Analogously, if payoffs are equal, we use a `$0$', and if the payoff is smaller, we use `$-$'. 
    Provided the benefit $b\!>\!1$, these relationships are independent of the exact value of $b$.
    }
    \label{tab:memory-1}
\end{table}

\noindent
{\bf Evolutionary dynamics in group-structured populations.}
To study the evolution of reciprocity in group-structured populations, we extend the models of Hauert, Chen, and Imhof~\cite{hauert2012evolutionary,hauert2014fixation} (who introduced the formalism for one-shot, non-repeated games).
We consider a finite population of size $MN$ subdivided into $M$ groups of size $N$.
Individuals in each group engage in pairwise interactions with all other members of their group, as depicted in Fig.~\ref{fig:model}{\color{blue}a}.

To compute the overall payoff of a focal individual with strategy $\mathbf{p}$, let $n_\mathbf{q}$ denote the number of group members that adopt strategy $\mathbf{q}\in\mathcal{M}$. Because the $\mathbf{p}$-player interacts with all $N\!-\!1$ other group members, the player's average payoff is 
\begin{equation}\label{eq:grouppayoff}
\pi_\mathbf{p} = \frac{1}{N-1}\sum_{\mathbf{q}\in\mathcal{M}} (n_\mathbf{q}-\delta_{\mathbf{p},\mathbf{q}}) \pi_{\mathbf{p},\mathbf{q}}
\end{equation}
Here, $\pi_{\mathbf{p},\mathbf{q}}$ is the payoff that the $\mathbf{p}$-player obtains against a co-player with strategy $\mathbf{q}$, as defined by Eq.~\eqref{eq:long_term_payoff}, and $\delta_{\mathbf{p},\mathbf{q}}$ is the indicator function that is one if $\mathbf{p}\!=\!\mathbf{q}$, and zero otherwise. 

The players' strategies are not fixed. Instead, each player updates its strategy according to the following dynamics.
At each time step, one player from the population is chosen randomly as the focal player. Suppose this player currently uses strategy $\mathbf{p}$. The focal player is then given a chance to adapt its strategy, either by intra-group imitation (with probability $\mu_{\rm in}$), out-group imitation (with probability $\mu_{\rm out}$), or mutation (with probability $\nu$), as depicted in Fig.~\ref{fig:model}(b-d). In particular, $\mu_{\rm in} + \mu_{\rm out} + \nu = 1$. 
In case of an intra-group imitation,the focal player randomly selects a role model from its own group. If the role model adopts strategy $\mathbf{q}$, the focal player switches to the role model's strategy with a probability given by the Fermi function\cite{blume:GEB:1993,szabo:PRE:1998}
\begin{equation} \label{Eq:f_in}
    f_{\mathbf{p} \to \mathbf{q}}^{\rm in} = \frac{1}{1+ \exp\left[\sigma_{\rm in} \left(\pi_\mathbf{p} - \pi_\mathbf{q} \right)\right]}. 
\end{equation}
Here, $\sigma_{\rm in}\!\ge\!0$ represents the selection strength of intra-group imitation. If $\sigma_{\rm in}$ is small, imitation events are mostly driven by chance,  $f_{\mathbf{p} \to \mathbf{q}}^{\rm in} \approx 1/2$. In contrast, if $\sigma_{\rm in}$ is large, the role model's strategy only has a reasonable chance of being imitated if it yields at least the payoff of the focal player. 

The case of out-group imitation follows an analogous procedure. Here, the focal player randomly selects a role model from a different group (with all other groups being equally likely). If the respective role model happens to use strategy $\mathbf{q}$, the focal player adopts this strategy with probability 
\begin{equation} \label{Eq:f_out}
f_{\mathbf{p} \to \mathbf{q}}^{\rm out} = \frac{1}{1+ \exp\left[\sigma_{\rm out} \left(\pi_\mathbf{p} - \pi_\mathbf{q} \right)\right]}.
\end{equation}
Here, $\sigma_{\rm out}$ is the selection strength for out-group imitation. Note that because the focal player and the role model are now in different groups, they do not play the game with each other, which is one of the key differences from models without group structure. 
Out-group imitation plays a similar role as migrations in genetic models of evolution~\cite{traulsen2006evolution,hauert2012evolutionary}. It allows strategies to move from one group to another. 
Finally, in case the focal player changes its strategy by mutation, the player simply replaces its current strategy $\mathbf{p}$ by a random strategy $\mathbf{q}$. All deterministic memory-1 strategies $\mathbf{q}$ are equally likely to be chosen. 

The above elementary updating process is iterated for many time steps. In each time step, a single individual is given the chance to update its strategy by intra-group imitation, out-group imitation, or mutation. Overall, this gives rise to a stochastic process on the space of all population compositions. In contrast to typical multilevel selection models~\cite{traulsen2006evolution,simon:Evolution:2012}, selection always operates on the individual level. Successful groups do not replace less successful ones; rather strategies of successful players are more likely to be imitated over time. 
The resulting stochastic process is straightforward to simulate. In the following, we derive results for two important special cases. In the first special case, we assume a complete separation of time-scales. Here, mutations are rare compared to out-group comparisons, which themselves are rare compared to intra-group comparisons, $\nu \ll \mu_{\rm out} \ll \mu_{\rm in}$. In the second case, we assume that mutations and out-group comparisons happen on a similar time scale, but both are rare compared to intra-group comparisons, $\nu\ll \mu_{\rm in}$ and $\mu_{\rm out} \ll \mu_{\rm in}$.

\section*{Analysis}

\subsection*{Dynamics when there is a full separation of time scales}

We begin by assuming a complete separation of time scales, $\nu \ll \mu_{\rm out} \ll \mu_{\rm in}$. 
In this setting, the intra-group dynamics are fast compared to the others. As a result, at any point in time there are at most two different strategies present in any group. 
When a mutation or an out-group imitation introduces a new strategy, intra-group imitation leads to the extinction or fixation of this strategy before the next strategy is introduced. In the following, we describe this dynamics in more detail.\\

\noindent
{\bf Description of the intra-group dynamics.} Consider a group in which $i$ players adopt the strategy $\mathbf{p}$ and $N\!-\!i$ players adopt the strategy $\mathbf{q}$. By Eq.~\eqref{eq:grouppayoff}, the players' payoffs are given by
\begin{equation}\label{eq:pi_xj}
\setlength{\arraycolsep}{1pt}
\begin{array}{lccc}
\displaystyle  \pi_\mathbf{p}(i) = &\displaystyle \frac{i-1}{N-1}\cdot \pi_{\mathbf{p},\mathbf{p}} &+ &\displaystyle \frac{N-i}{N-1}\cdot \pi_{\mathbf{p},\mathbf{q}},\\[0.4cm]
\displaystyle  \pi_\mathbf{q}(i) = &\displaystyle \frac{i}{N-1}\cdot \pi_{\mathbf{q},\mathbf{p}} &+ &\displaystyle \frac{N-i-1}{N-1}\cdot \pi_{\mathbf{q},\mathbf{q}}. 
  \end{array}
\end{equation}
The probability that intra-group imitation increases (decreases) the number of $\mathbf{p}$-players in a single time step is
\begin{align}
T_i^{+} &= \frac{N-i}{N}\frac{i}{N-1} \frac{1}{1+ \exp\left\{\sigma_{\rm in} \left[\pi_\mathbf{q}(i) - \pi_\mathbf{p}(i) \right]\right\}} \\[0.2cm]
T_i^{-} &= \frac{N-i}{N}\frac{i}{N-1} \frac{1}{1+ \exp\left\{\sigma_{\rm in} \left[\pi_\mathbf{p}(i) - \pi_\mathbf{q}(i) \right]\right\}}.
\end{align}
Here, the first equation corresponds to the case where a $\mathbf{q}$-player is randomly chosen as the updating player, a $\mathbf{p}$-player is chosen as the role model, and the updating player chooses to imitate the role model. The second equation corresponds to the converse case of a $\mathbf{p}$-player imitating a role model with strategy $\mathbf{q}$. 

The fixation probability of a single $\mathbf{p}$ player in a resident group of $\mathbf{q}$-players can be computed explicitly\cite{nowak:Nature:2004,traulsen:bookchapter:2009}. This probability is given by $\rho_{\mathbf{p},\mathbf{q}} = \big( 1 + \sum_{j=1}^{N-1} \prod_{i=1}^j \gamma_i \big)^{-1}$, where $\gamma_i \equiv T_i^{-} / T_i^{+} = \exp\left[\sigma_{\rm in}\cdot \left( \pi_\mathbf{q}(i) - \pi_\mathbf{p}(i) \right) \right]$.
By using the explicit payoff equations~(\ref{eq:pi_xj}), this fixation probability becomes~\cite{stewart2013extortion}
\begin{equation}\label{eq:intra_fixation_prob}
\rho_{\mathbf{p},\mathbf{q}} = \left( \sum_{i=0}^{N-1} \exp\left[ \sigma_{\rm in} \cdot i \frac{ (2N-i-3)\pi_{\mathbf{q},\mathbf{q}} + (i+1)\pi_{\mathbf{q},\mathbf{p}} - (2N-i-1)\pi_{\mathbf{p},\mathbf{q}} - (i-1)\pi_{\mathbf{p},\mathbf{p}}}{2(N-1)} \right] \right)^{-1}.
\end{equation}
Using this formula, we can compute for each resident strategy $\mathbf{q}$ how likely it is that any novel strategy $\mathbf{p}$ is eventually adopted by the entire group. 
While the use of fixation probabilities has become common practice in evolutionary game theory\cite{traulsen:bookchapter:2009}, we note that the time it takes for a single strategy to reach fixation may be considerable. The fixation time becomes particularly long when groups are large, and when the strategies $\mathbf{p}$ and $\mathbf{q}$ allow for an equilibrium in which the two strategies stably co-exist\cite{wu:JMB:2012}. Nevertheless, this limit has become a useful approximation, as it simplifies computations considerably. Instead of considering arbitrarily many strategies at once, one can make predictions by only considering two strategies at a time\cite{fudenberg:JET:2006,mcavoy:jet:2015}. Once a strict separation of time scales does no longer apply, the analysis becomes considerably more intricate\cite{vasconcelos:prl:2017}.\\


\noindent
{\bf Description of the inter-group dynamics.}
To further simplify the analysis of our model, we make the additional assumption that $\nu \ll \mu_{\rm out}$.
This limit indicates that the time scale for out-group imitations is short compared to the time scale of mutations.
This assumption implies that at any point in time, at most two different strategies are present in the entire population.
Once a mutation introduces a new strategy, this strategy either fixes in the population (through successive in-group and out-group imitation events), or the strategy goes extinct. 
To describe this dynamics in more detail, suppose that the two strategies $\mathbf{p}$ and $\mathbf{q}$ are present in the population. 
Since intra-group imitation is fast, every group is homogeneous. As a consequence, we can speak of $\mathbf{p}$-groups and $\mathbf{q}$-groups, depending on which strategy the group members employ. Once a $\mathbf{q}$-player imitates a player from a $\mathbf{p}$-group, the number of $\mathbf{p}$-groups may increase (if the strategy $\mathbf{p}$ reaches fixation in the $\mathbf{q}$-group). 
The respective probability that the number $i$ of $\mathbf{p}$-groups increases (or decreases) is given by
\begin{align}
Q_i^{+} &=  \frac{i}{M}\frac{M-i}{M-1} \frac{1}{1+ \exp\left[\sigma_{\rm out} \left(\pi_{\mathbf{q},\mathbf{q}} - \pi_{\mathbf{p},\mathbf{p}} \right)\right]} \cdot \rho_{\mathbf{p},\mathbf{q}}, \\[0.2cm]
Q_i^{-} &=  \frac{i}{M}\frac{M-i}{M-1} \frac{1}{1+ \exp\left[\sigma_{\rm out} \left(\pi_{\mathbf{p},\mathbf{p}} - \pi_{\mathbf{q},\mathbf{q}} \right)\right]} \cdot \rho_{\mathbf{q},\mathbf{p}},
\end{align}
respectively. In both expressions, the first three factors on the right hand side represent the probability of the respective out-group imitation event. The last factor is the probability that the newly introduced strategy reaches fixation. 
The ratio $\eta$ of these transition probabilities simplifies to
\begin{equation}\label{eq:eta}
\eta \equiv \frac{Q_i^{-}}{Q_i^{+}} = \frac{\rho_{\mathbf{q},\mathbf{p}}}{\rho_{\mathbf{p},\mathbf{q}}} \exp\left[ \sigma_{\rm out} (\pi_{\mathbf{q},\mathbf{q}} - \pi_{\mathbf{p},\mathbf{p}}) \right].
\end{equation}
From Eq.~(\ref{eq:intra_fixation_prob}), the ratio of the intra-group fixation probabilities is~\cite{traulsen2008analytical}
\begin{equation}\label{eq:intra_fixation_prob_ratio}
    \frac{\rho_{\mathbf{p},\mathbf{q}}}{\rho_{\mathbf{q},\mathbf{p}}} = \prod_{j=1}^{N-1} \gamma_j = \exp\left\{ \sigma_{\rm in} \left[ \Big(\pi_{\mathbf{p},\mathbf{p}} - \pi_{\mathbf{q},\mathbf{q}}\Big) \left(\frac{N}{2}-1\right) + \Big(\pi_{\mathbf{p},\mathbf{q}} - \pi_{\mathbf{q},\mathbf{p}}\Big) \frac{N}{2} \right] \right\}.
\end{equation}
Thus, Eq.~(\ref{eq:eta}) can be re-written as
\begin{equation}\label{eq:eta_general_N}
\eta = \exp\left\{ -\sigma_{\rm in} \left[ \Big(\pi_{\mathbf{p},\mathbf{p}} - \pi_{\mathbf{q},\mathbf{q}} \Big) \left(\frac{N}{2}-1\right) + \Big(\pi_{\mathbf{p},\mathbf{q}} - \pi_{\mathbf{q},\mathbf{p}}\Big) \frac{N}{2} \right] + \sigma_{\rm out} \Big(\pi_{\mathbf{q},\mathbf{q}}-\pi_{\mathbf{p},\mathbf{p}}\Big) \right\}.
\end{equation}

\noindent
Overall, we obtain the following formula for the probability that a new strategy $\mathbf{p}$ takes over the entire population, given everyone else applies strategy $\mathbf{q}$, 
\begin{equation}\label{eq:psi}
\Psi_{\mathbf{p},\mathbf{q}} = \rho_{\mathbf{p},\mathbf{q}} \frac{1}{1 + \sum_{j=1}^{M-1} \eta^{j} } =
        \begin{cases}
          \rho_{\mathbf{p},\mathbf{q}} \frac{1-\eta}{ 1-\eta^M } & \text{when $\eta \neq 1$} \\[0.15cm]
          \rho_{\mathbf{p},\mathbf{q}} / M                       & \text{when $\eta = 1$,}
       \end{cases}
\end{equation}
Here, the first factor $\rho_{\mathbf{p},\mathbf{q}}$ is the probability that the $\mathbf{p}$-mutant takes over the first group. 
The second factor gives the probability that eventually also all other groups adopt $\mathbf{p}$. 
Similarly one can calculate the fixation probability that everyone in a $\mathbf{p}$-population eventually imitates a single $\mathbf{q}$ mutant. This probability is 
\begin{equation} \label{eq:psi2}
\Psi_{\mathbf{q},\mathbf{p}} = \rho_{\mathbf{q},\mathbf{p}} \frac{1}{1 + \sum_{j=1}^{M-1}\eta^{-j} } =
       \begin{cases}
          \rho_{\mathbf{q},\mathbf{p}} \frac{1-\eta^{-1}}{ 1-\eta^{-M} } & \text{when $\eta \neq 1$} \\[0.15cm]
          \rho_{\mathbf{q},\mathbf{p}} / M                       & \text{when $\eta = 1$.}
       \end{cases}
\end{equation}
These formula allow us to compute how likely any given mutant strategy is to replace the resident strategy when there is a complete separation of time scales.\\

\noindent
{\bf Strategies favored by the evolutionary process.}
Using these formulas, we can analyze which strategies are particularly likely to spread. To this end, we say strategy $\mathbf{p}$ is favored over $\mathbf{q}$ if a single $\mathbf{p}$-mutant is more likely to fix in a $\mathbf{q}$-population than vice versa.  By Eqs.~\eqref{eq:psi} and~\eqref{eq:psi2}, the respective condition $\Psi_{\mathbf{p},\mathbf{q}} > \Psi_{\mathbf{q},\mathbf{p}}$ simplifies to
\begin{equation}\label{eq:X_win}
\frac{\rho_{\mathbf{p},\mathbf{q}}}{\rho_{\mathbf{q},\mathbf{p}}} >  \exp\left[ \frac{M-1}{M} \sigma_{\rm out} \Big(\pi_{\mathbf{q},\mathbf{q}}-\pi_{\mathbf{p},\mathbf{p}} \Big) \right].
\end{equation}
The left hand side reflects the effect of in-group imitation, whereas the right hand side captures the effect of out-group imitation. 
In the special case of a single group, $M=1$, this condition reproduces the respective condition for well-mixed populations, $\rho_{\mathbf{p},\mathbf{q}} > \rho_{\mathbf{q},\mathbf{p}}$.
Plugging Eq.~(\ref{eq:intra_fixation_prob_ratio}) into Eq.~(\ref{eq:X_win}) yields
\begin{equation}\label{eq:x_vs_y}
\sigma_{\rm in} \frac{N}{2} \pi_{\mathbf{p},\mathbf{q}} + \left[ \sigma_{\rm in} \left( \frac{N}{2}-1 \right) + \sigma_{\rm out} \frac{M-1}{M} \right] \pi_{\mathbf{p},\mathbf{p}} > \sigma_{\rm in} \frac{N}{2} \pi_{\mathbf{q},\mathbf{p}} + \left[ \sigma_{\rm in} \left( \frac{N}{2} - 1 \right) + \sigma_{\rm out} \frac{M-1}{M} \right] \pi_{\mathbf{q},\mathbf{q}}.
\end{equation}
By collecting alike terms, this expression can be further simplified to 
\begin{equation} \label{Eq:pfavored}
\sigma_{\rm in} \left[ \frac{N}{2} \big(\pi_{\mathbf{p},\mathbf{q}} -\pi_{\mathbf{q},\mathbf{p}} \big) + \left( \frac{N}{2}-1 \right) \big( \pi_{\mathbf{p},\mathbf{p}} - \pi_{\mathbf{q},\mathbf{q}} \big) \right] + \sigma_{\rm out} \frac{M-1}{M}  \big( \pi_{\mathbf{p},\mathbf{p}} - \pi_{\mathbf{q},\mathbf{q}} \big) > 0.
\end{equation}
The first and the second terms of this inequality correspond to the dynamics within and between groups, respectively. 
The intra-group dynamics is decisive if either $\sigma_{\rm in}N \gg \sigma_{\rm out}$ or if the number of groups is small ($M\approx 1$). In that case, and if groups are additionally assumed to be small ($N\rightarrow2$), the condition for $\mathbf{p}$ to be favored simplifies to 
\begin{equation}
 \pi_{\mathbf{p},\mathbf{q}} > \pi_{\mathbf{q},\mathbf{p}}.
 \end{equation}
 This condition is closely related to the notion of rival strategies\cite{hilbe2015partners}. Strategy $\mathbf{p}$ is a rival strategy if and only if it enforces the condition $ \pi_{\mathbf{p},\mathbf{q}} \ge \pi_{\mathbf{q},\mathbf{p}}$ against all strategies $\mathbf{q}$. In Table~\ref{tab:memory-1}, the second-to-last column indicates all memory-1 rival strategies. There are four of them, TFT, Grim, AllD, and the strategy $\mathbf{p}=(0,0,1,0)$. The above observations suggest that these rival strategies should be particularly strong when there is only a single group with two group members ($M=1$ and $N=2$). 

In the other extreme, when $\sigma_{\rm in}N \ll \sigma_{\rm out}$ and $M$ is sufficiently large, it is the inter-group dynamics that is decisive. In that case, the relative strength of a strategy is determined by its efficiency. Strategy $\mathbf{p}$ is favored over $\mathbf{q}$ if and only if it yields the larger payoff against itself, 
\begin{equation}
 \pi_{\mathbf{p},\mathbf{p}} > \pi_{\mathbf{q},\mathbf{q}}.
 \end{equation}
A final interesting case arises when the two selection strength parameters are equal, $\sigma_{\rm in} = \sigma_{\rm out}$. In that case, condition~\eqref{Eq:pfavored} simplifies to 
\begin{equation}
MN\big(\pi_{\mathbf{p},\mathbf{q}} -\pi_{\mathbf{q},\mathbf{p}} \big) + (MN-2) \big( \pi_{\mathbf{p},\mathbf{p}} - \pi_{\mathbf{q},\mathbf{q}} \big)  > 0.
\end{equation}
In particular, if the total population becomes large $MN\rightarrow \infty$, strategy $\mathbf{p}$ is favored if and only if 
\begin{equation} \label{eq:x_vs_y_simplified}
\pi_{\mathbf{p},\mathbf{p}} +\pi_{\mathbf{p},\mathbf{q}} - \pi_{\mathbf{q},\mathbf{p}} - \pi_{\mathbf{q},\mathbf{q}}  > 0.
\end{equation}
That is, $\mathbf{p}$ is favored if and only if it is risk-dominant\cite{harsanyi1988general}, independent of the exact values of $M$ and $N$.\\

\noindent
{\bf Numerical simulations.} 
The above arguments are valid only when players choose among two strategies.
In the following, we explore evolution among all 16 deterministic memory-1 strategies by implementing the evolutionary process numerically. To this end, we use Monte Carlo simulations. Mutant strategies are repeatedly introduced into the current resident population. The mutant strategy either takes over or goes extinct. We report how much cooperation we observe on average (see Methods). 

\begin{figure}[t]
    \centering
    \includegraphics[width=0.9\textwidth]{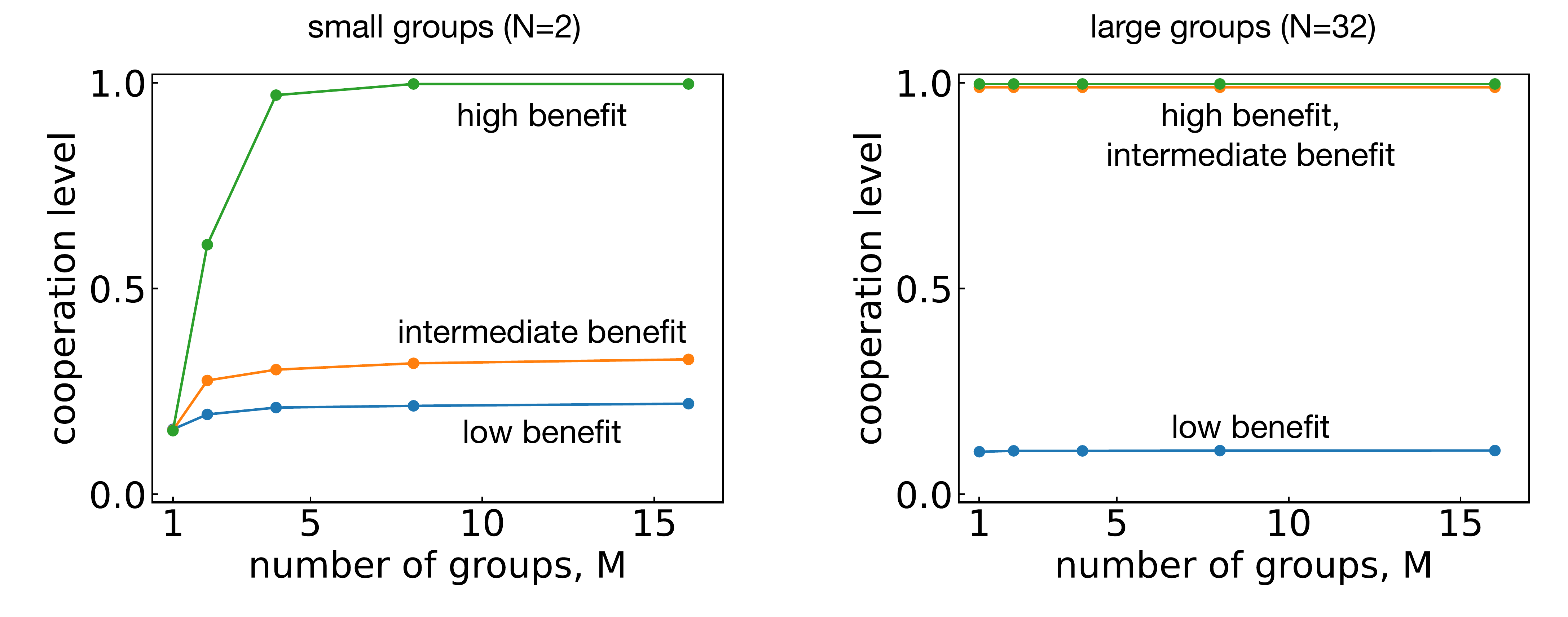}
    \caption{
    Cooperation levels as functions of the number of groups $M$ for small groups ($N=2$) and large groups ($M=32$).
    The curves corresponds to high benefit ($b=6$), intermediate benefit ($b=3$), and low benefit ($b=1.5$) from top to bottom.
    When groups are small, we see more cooperation when we increase the number of groups. In contrast, when each group is large, the exact number of groups has little effect on the average cooperation level.}
    \label{fig:coop_M_dep_N_fixed}
\end{figure}

Figure~\ref{fig:coop_M_dep_N_fixed} shows how the evolving cooperation level depends on the number of groups $M$, either for small groups ($N=2$) or for relatively large groups ($N=32$). 
When the group size is small, we observe very little cooperation if there is only a single group ($M=1$), as predicted by our earlier analysis. As we increase the number of groups, also the cooperation level increases. However, they do not improve indefinitely. Rather, these improvements saturate as we increase $M$, which is consistent with the factor $(M-1)/M$ in Eq.~(\ref{eq:x_vs_y}).
The limiting cooperation level depends on the benefit of cooperation, reproducing the standard result that larger benefits are more conducive to cooperation\cite{hilbe2018partners}. 
In general, we thus observe that cooperation tends to be favored when $M$, $N$, and $b$ are large, corresponding to many groups of substantial size, and a considerable benefit to cooperation. 

\begin{figure}[t!]
    \centering
    \includegraphics[width=0.9\textwidth]{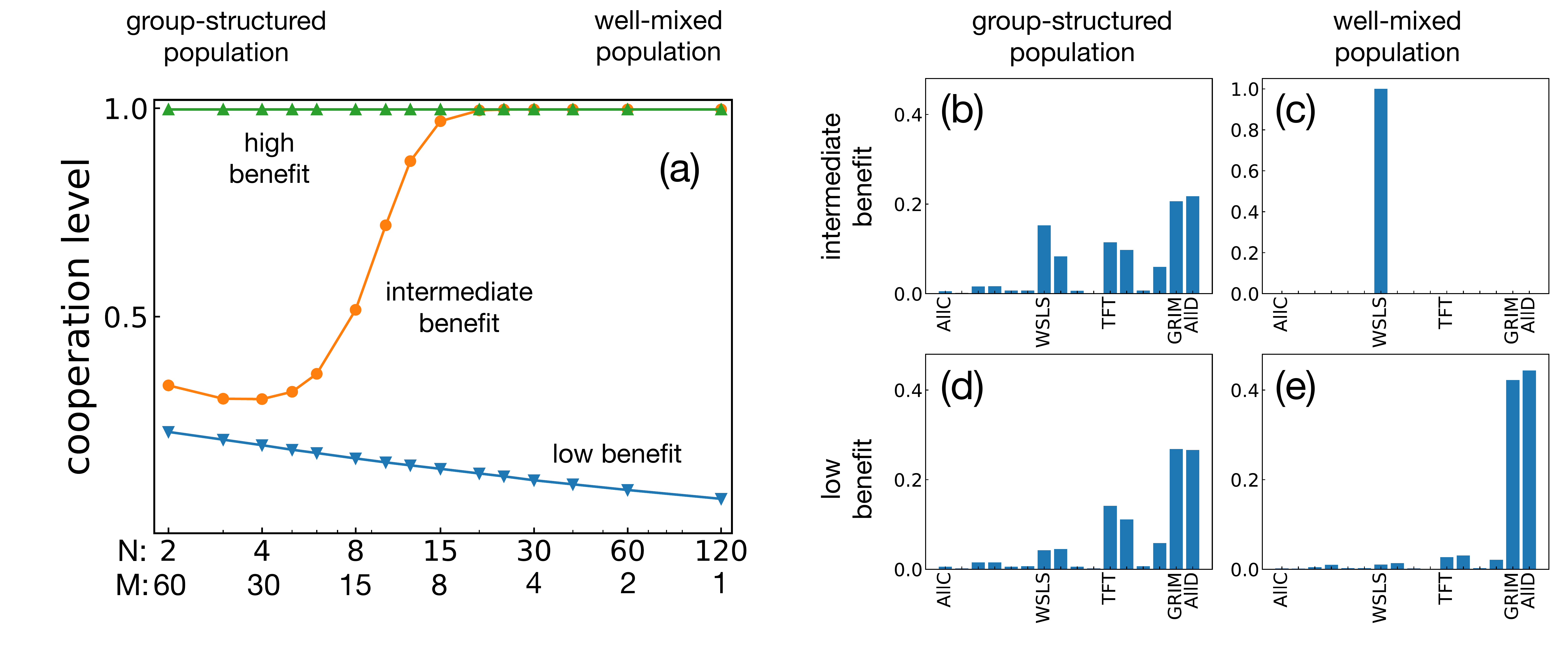}
    \caption{The effect of group structure on the evolution of cooperation. Here we keep the total population size fixed to $MN=120$, while simultaneously varying the size $N$ of each group and the number of groups $M=120/N$. 
    (a) We observe that group structure has a positive effect for small $b$, a negative effect for intermediate $b$ and no effect for large $b$. 
    (b-e) To explore these non-trivial effects of group structure in more detail, we record the abundance of each strategy for four different scenarios. The scenarios differ in whether the benefit is low or intermediate ($b=1.5$ and $b=3$, respectively), and whether the population is group-structured or well-mixed ($N\!=\!2$, $M\!=\!60$ and $N\!=\!120$, $M\!=\!1$, respectively). Note that the different panels use different scales for the vertical axes.}
    \label{fig:low_mut_abundance}
\end{figure}

After exploring the effect of group size and number of groups in isolation, we next ask to which extent group structure facilitates cooperation. 
To this end, we keep the total population size fixed at $MN=120$, and vary the group size $N$. The number of groups is then automatically determined as $M=120/N$. In one extreme case, there is only a single group of maximum size, $M=1$ and $N=120$. We refer to this scenario as the case of a (single) well-mixed population. In the other extreme case, groups take the minimum non-trivial size, $N=2$, which implies that the resulting number of groups is $M=60$. We refer to this second scenario as the case of a (fully) group-structured population. 
Figure~\ref{fig:low_mut_abundance}(a) shows how the cooperation level changes as we vary the group size $N$.
Interestingly, the effect of group structure depends on the benefit~$b$ of cooperation. 
If the benefit is small, group-structured populations achieve more cooperation than well-mixed populations. 
For intermediate benefits, we observe the opposite trend. Here, well-mixed populations are more conducive to cooperation. 
Finally, once benefits are very large, full cooperation evolves in all considered cases, independent of the exact values of $M$ and $N$. 

To further investigate these non-trivial effects of group structure, we analyze the abundance of each of the 16 strategies in the stationary state, see Figure~\ref{fig:low_mut_abundance}(b-e). 
We first consider the case that the benefit of cooperation is intermediate, $b\!=\!3$. Here, well-mixed populations lead to much more cooperation. In particular, here we observe that populations learn to adopt the cooperative Win-Stay Lose-Shift (WSLS) strategy almost all of the time, Figure~\ref{fig:low_mut_abundance}(c). 
In group-structured populations, on the other hand, no single strategy is predominant. The most abundant strategies are the non-cooperative strategies AllD and Grim. The next abundant strategies are WSLS and TFT, respectively. 
Overall, we thus observe that well-mixed populations are more favorable for cooperation because they make it more likely that the cooperative strategy WSLS evolves. 
In a second step, we consider the case of low cooperation benefits, see Figure~\ref{fig:low_mut_abundance}(d,e). Group-structured populations again lead to the evolution of the strategies AllD, Grim, TFT, and WSLS (and related strategies). In contrast, well-mixed populations consist of the non-cooperative strategies AllD and Grim almost entirely. 

\begin{figure}[t]
    \centering
    \includegraphics[width=0.9\textwidth]{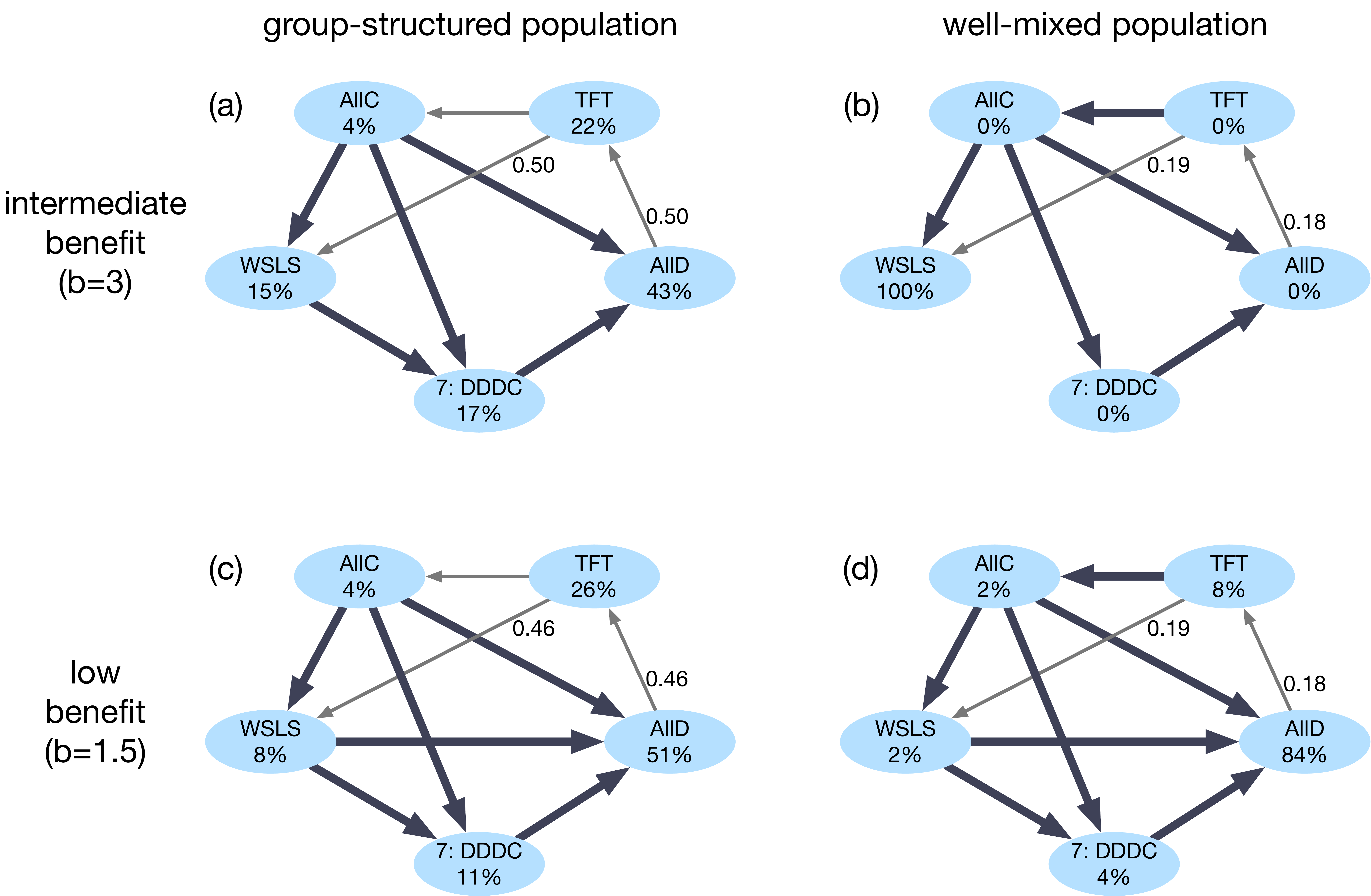}
    \caption{Evolutionary dynamics among five representative strategies when there is a complete separation of time scales. 
    To shed further light on the previous evolutionary results among all memory-1 strategies, here we display the transitions for a reduced strategy set. 
    The reduced strategy set consists of the five strategies AllD, TFT, AllC, WSLS, and $S_7=(0,0,0,1)$. 
    We consider four scenarios, depending on whether the benefit of cooperation is intermediate or low (top and bottom), and depending on whether the population is group-structured or well-mixed (left and right, respectively). Group-structured populations consist of $M=60$ groups of size $N=2$; well-mixed populations consist of a single $M=1$ group of size $N=120$. 
    The thickness of the arrows indicates the respective fixation probabilities.
    Thick arrows without numbers represent cases in which the fixation probability is approximately one. 
    For better clarity, we only depict arrows when the fixation probability is at least 0.1. The numbers represent numerically exact values for a selection strength of $\sigma_{\rm in} = \sigma_{\rm out} = 10$. 
    }
    \label{fig:low_mut_transition}
\end{figure}

To better understand why cooperative strategies are abundant in one scenario but not in another, we investigate the transition probabilities for a reduced strategy space. 
The reduced strategy space contains the representative strategies AllC, WSLS, TFT, AllD and the strategy $S_7$ with $\mathbf{p}=(0,0,0,1)$. 
The payoffs and the win-lose relationships of these strategies are summarized in Tables~\ref{tab:payoffs_major_strategies} and~\ref{tab:signs_major_strategies}.
In addition, Figure~\ref{fig:low_mut_transition} illustrates the average abundance of each of these five strategies and the transition probabilities between them.
We confirmed that the overall cooperation levels for this reduced strategy space are comparable to the corresponding results for the full strategy space. Hence, the reduced strategy space can serve a useful proxy to gain insights into the overall dynamics.

\begin{table}[t]
    \centering
    \begin{tabular}{|c|c|c|c|c|c|}
    \hline
               & AllC     & WSLS       &$S_7$  & TFT        & AllD    \\ \hline
    AllC       & $b-1$    & $1/2b-1$   & $-1$        & $b-1$      & $-1$    \\
    WSLS       & $b-1/2$  & $b-1$      & $1/3b-2/3$  & $1/2b-1/2$ & $-1/2$  \\
    $S_7$ & $b$      & $2/3b-1/3$ & $1/2b-1/2$  & $1/3b-1/3$ & $-1/2$  \\
    TFT        & $b-1$    & $1/2b-1/2$ & $1/3b-1/3$  & $1/2b-1/2$ & $0$     \\
    AllD       & $b$      & $1/2b$     & $1/2b$      & $0$        & $0$     \\
    \hline
    \end{tabular}
    \caption{Payoff matrix for the reduced strategy space, consisting of AllC, WSLS, $S_7$, TFT, and AllD. 
    The entry $(i,j)$ is the payoff $\pi_{ij}$ that the (row) strategy $i$ obtains against the (column) strategy $j$.\newline}
    \label{tab:payoffs_major_strategies}

    \centering
    \begin{tabular}{|c|c|c|c|c|c|}
    \hline
               & AllC     & WSLS       &$S_7$  & TFT        & AllD    \\ \hline
    AllC       &          & $-$        & $-$         & $+$        & $-$    \\
    WSLS       & $+$      &            & $b > 5$     & $+$        & $b > 3$  \\
    $S_7$ & $+$      & $b < 5$    &             &            & $-$  \\
    TFT        & $-1$     & $-$        &             &            & $+$     \\
    AllD       & $+$      & $b < 3$    & $+$         & $-$        &         \\
    \hline
    \end{tabular}
    \caption{Pairwise comparisons for the reduced strategy space. Again, we consider every pairwise combination of AllC, WSLS, $S_7$, TFT, and AllD. The cells indicate the sign of $\pi_{\mathbf{p},\mathbf{p}} + \pi_{\mathbf{p},\mathbf{q}} - \pi_{\mathbf{q},\mathbf{p}} - \pi_{\mathbf{q},\mathbf{q}}$, the left-hand side of Eq.~\eqref{eq:x_vs_y_simplified}.
A plus sign (minus sign) indicates that the row strategy $\mathbf{p}$ is favored (disfavored) over the column strategy $\mathbf{q}$. 
    If the sign depends on the benefit $b$, the range for which the sign is positive is shown.
    Blank cells indicate that the left-hand side of Eq.~\eqref{eq:x_vs_y_simplified} is equal to zero. 
    }
    \label{tab:signs_major_strategies}
\end{table}

We first consider the case of an intermediate benefit, $b\!=\!3$. Here, well-mixed populations yield more cooperation, as they promote the evolution of WSLS.  Fig.~\ref{fig:low_mut_transition}(b) shows why. 
There is an evolutionary path from every other strategy towards WSLS; once the entire population adopts WSLS, every other mutant strategy is at a disadvantage. This picture is in line with previous research on direct reciprocity in well-mixed populations\cite{baek2016comparing}. 
The picture changes, however, in group-structured populations, see Fig.~\ref{fig:low_mut_transition}(a) depicting the case of groups of size $N\!=\!2$. Here, a homogeneous WSLS population can be invaded by $S_7$. To see why, consider a group that contains both strategies. By Table~\ref{tab:payoffs_major_strategies}, the payoff of WSLS is $1/3b-2/3$, which is below the payoff of $S_7$, $2/3b-1/3$. Hence, $S_7$ is favored in each mixed group. On the other hand, with respect to out-group imitation, it is WSLS that is favored over $S_7$, because the payoff of WSLS against itself is $b\!-\!1$, which exceeds $S_7$'s self-payoff of $(b-1)/2$. To compute which of the two opposing effects dominates, Eq.~\eqref{Eq:pfavored}  suggests that we need to compute the sign of $\pi_{\mathbf{p},\mathbf{p}} +\pi_{\mathbf{p},\mathbf{q}} - \pi_{\mathbf{q},\mathbf{p}} - \pi_{\mathbf{q},\mathbf{q}}$. For $b<5$, this criterion suggests that $S_7$ is favored (as also indicated in Table~\ref{tab:signs_major_strategies}). These observations explain why in group-structured populations, WSLS is susceptible to invasion by $S_7$, which in turn can be invaded by AllD. 

In a next step, we explore the case of a small benefit of cooperation, $b\!=\!1.5$. Here, group-structured populations are more cooperative. 
The respective transition graphs for group-structured and well-mixed populations are depicted in Figs.~\ref{fig:low_mut_transition}(c) and (d).
In both cases, we observe that there is no single strategy that resists invasion by all other strategies. 
Instead, AllD populations are susceptible to TFT, which in turn is susceptible to AllC and WSLS, which can be invaded by AllD again. 
The main difference between group-structured and well-mixed populations is the relative performance of TFT. 
Compared to well-mixed populations, TFT is better able to invade AllD populations in structured populations. To see why, we first consider the within-group dynamics when $N=2$. Because TFT gets the same payoff as the opponent in any pairwise encounter\cite{press2012iterated}, the fixation probability of TFT in a group with ALLD is exactly 1/2. In addition, TFT is favored by the between-group dynamics, because the payoff of TFT-groups is (b-c)/2, which is larger than the payoff of zero in AllD-groups. It follows that a single TFT mutant can replace an AllD population with a probability that is approximately 1/2. In contrast, in well-mixed populations, this fixation probability is much smaller, 0.18 for the parameters in Fig.~\ref{fig:low_mut_transition}(d). 

The above results suggest that overall, there are two competing effects when splitting a population into smaller groups.
On the one hand, smaller group sizes favor the evolution of rival strategies because small groups generally select for spite\cite{nowak:Nature:2004}.
On the other hand, group structure can favor the evolution of cooperation because individuals in highly cooperative groups are more likely imitated. Our above results suggest that the overall outcome of these two opposing effects depends on the benefit of cooperation. When this benefit is comparably small, group-structured populations allow for more cooperation than well-mixed populations. In contrast, when this benefit is intermediate, cooperation in well-mixed populations is more robust.

\subsection*{Dynamics when there is a partial separation of time scales}

Throughout our analysis so far we have assumed a complete separation of time-scales. When a player was randomly chosen to update its strategy, we assumed that this player is most likely to engage in intra-group imitation, far less likely to engage in out-group imitation, and again far less likely in random exploration (mutation). In the following, we instead assume that intra-group comparisons are still most likely; however, mutations and out-group comparisons now occur on a similar time scale. In this limit, all groups can be assumed to be homogeneous because intra-group imitation is fast. However, different groups might employ different strategies, because mutations might introduce novel strategies faster than out-group imitation can result in the fixation of any given strategy in the population.\\

\noindent
{\bf A differential equation in the limit of large populations.} To obtain analytical results, in the following we assume that the number of groups is large, $M\rightarrow \infty$. Let $x_\mathbf{p}$ be the fraction of groups that employ strategy $\mathbf{p}$. Over time, these fractions can change, either because new strategies are introduced into groups by out-group imitation (and reach fixation), or they are introduced by mutations (and reach fixation). This dynamics may be described by the following differential equation, 
\begin{equation}\label{eq:finite_mut_theory}
\dot{x}_\mathbf{p} = \left(1- r \right)\sum_{\mathbf{q}\neq \mathbf{p}} \alpha_{\mathbf{p},\mathbf{q}} x_\mathbf{p} x_\mathbf{q}
  + r\sum_{\mathbf{q}\neq \mathbf{p}}\frac{x_\mathbf{q}\cdot \rho_{\mathbf{p},\mathbf{q}} - x_\mathbf{p}\cdot \rho_{\mathbf{q},\mathbf{p}}}{|\mathcal{M}|}.
\end{equation}
Here, $r=\nu/(\nu+\mu_{\rm out})$ is the relative mutation probability (compared to out-group imitation events). 
The right hand side of Eq.~\eqref{eq:finite_mut_theory} consists of two parts. The first sum describes the changes triggered by out-group imitation. Here, the parameter
\begin{equation}
    \alpha_{\mathbf{p},\mathbf{q}} \equiv \frac{\rho_{\mathbf{p},\mathbf{q}}}{ 1 + \exp\left[ \sigma_{\rm out} \left( \pi_{\mathbf{q},\mathbf{q}} - \pi_{\mathbf{p},\mathbf{p}} \right) \right] }
                     - \frac{\rho_{\mathbf{q},\mathbf{p}}}{ 1 + \exp\left[ \sigma_{\rm out} \left( \pi_{\mathbf{p},\mathbf{p}} - \pi_{\mathbf{q},\mathbf{q}} \right) \right] }
\end{equation}
describe the flow from strategy $\mathbf{q}$ to strategy $\mathbf{p}$. For example, the denominator of the first term on the right hand side describes the likelihood that a $\mathbf{q}$-player switches to $\mathbf{p}$ due to out-group imitation. The numerator describes the likelihood that subsequently, $\mathbf{p}$ reaches fixation due to in-group imitation. The interpretation of the second term on the right hand side is similar, by considering the possibility that a $\mathbf{p}$-group makes the converse transition towards $\mathbf{q}$. 
The second sum in Eq.~\eqref{eq:finite_mut_theory} describes the changes triggered by mutation events. Here, the denominator of $\rho_{\mathbf{p},\mathbf{q}}/|\mathcal{M}|$ describes the probability that the mutating player adopts strategy $\mathbf{p}$. The numerator gives the probability that this strategy is then adopted by all other group members, due to intra-group imitation. 
We note that the sum $\sum_\mathbf{p} x_\mathbf{p}=1$ by definition, and hence the equation is defined on the $16$-dimensional simplex. Moreover, since $\sum_\mathbf{p}{\dot{x}_\mathbf{p}} = 0$, the unit simplex is invariant under the dynamics. One may interpret Eq.~\eqref{eq:finite_mut_theory} as a variant of the replicator-mutator equation\cite{nowak:book:2006}, where the first part represents selection and the second part represents mutations.  

Further below, we explore the solutions of Eq.~\eqref{eq:finite_mut_theory} numerically, for various parameter combinations. 
For all parameters we considered, the dynamics converges to a stable fixed point.
Such a fixed point satisfies the equation
\begin{equation}
x_\mathbf{p}^{\ast} = \frac{\displaystyle
r\cdot \sum_{\mathbf{q}\neq \mathbf{p}}\rho_{\mathbf{p},\mathbf{q}} \, x_\mathbf{q}^{\ast}/|\mathcal{M}| }
{\displaystyle
 (1-r) \sum_{\mathbf{q}\neq \mathbf{p}} \alpha_{\mathbf{p},\mathbf{q}} x_\mathbf{q}^{\ast} - r\sum_{\mathbf{q}\neq \mathbf{p}}\rho_{\mathbf{q},\mathbf{p}}/|\mathcal{M}|}.
\end{equation}
We would like to emphasize that the Eq.~\eqref{eq:finite_mut_theory} does not need to recover the qualitative dynamics that we obtained in the previous section, even when $r\rightarrow 0$ (in which case mutations are again rare compared to out-group imitation events). 
In other words, the order in which limits are taken affects the solution that is predicted. 
As we show further below, however, the solutions predicted by Eq.~\eqref{eq:finite_mut_theory} are in excellent agreement with explicit simulations of the evolutionary process for all values of $r$ we considered.\\

\begin{figure}[t]
    \centering
    \includegraphics[width=0.9\textwidth]{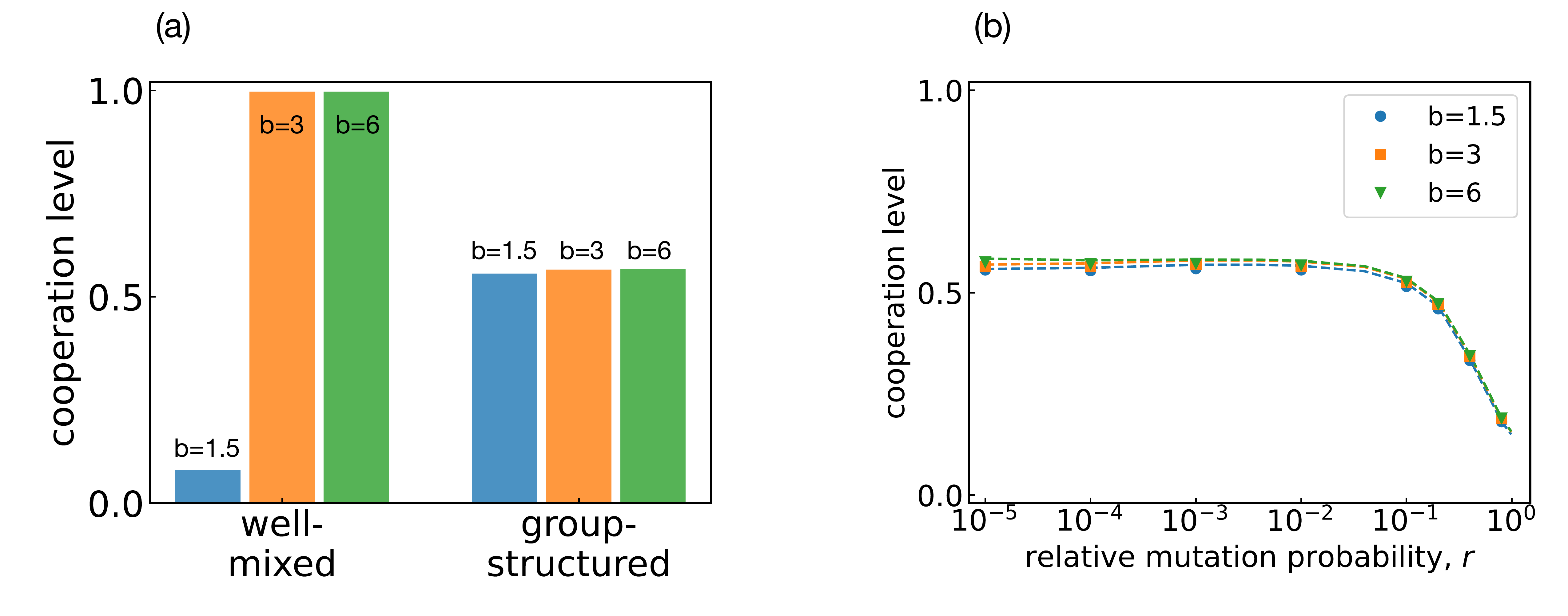}
    \caption{Comparison of well-mixed and group-structured populations when there is a partial separation of time scales. 
    (a) We consider a well-mixed population ($N=200$, $M=1$) and a group-structured population ($N=2$, $M=100$), assuming that the relative probability of mutations is $r=0.01$. In both cases we simulate the evolutionary process with Monte Carlo simulations and record how often players cooperate on average. For well-mixed populations, these average cooperation levels strongly depend on the benefit of cooperation. In contrast, for group-structured populations, average cooperation levels are approximately 1/2, independent of the considered cooperation benefit. In particular, we recover our previous result that group-structured populations yield more cooperation when $b$ is small, and less cooperation when $b$ is large. 
    (b)  The results are qualitatively robust across many values of $r$ (relative mutation probability). 
    The points in this graph depict the result of Monte Carlo simulations. Dashed lines are the predictions by Eq.~(\ref{eq:finite_mut_theory}).}
    \label{fig:finite_mut_c_level}
\end{figure}

\noindent
{\bf Numerical results.} 
Fig.~\ref{fig:finite_mut_c_level}(a) shows the evolving cooperation levels for a well-mixed population ($N\!=\!200, M\!=\!1$) and a group-structured population ($N\!=\!2$, $M\!=\!100$). We observe a striking difference between the two settings. 
In the well-mixed population, the cooperation level strongly depends on the benefit of cooperation, as one may expect. 
For small benefit values, hardly any cooperation evolves. For intermediate and large benefit values, almost the entire population cooperates eventually. 
In contrast, in the group-structured population, cooperation levels are around $1/2$ when $r$ is low, largely independent of the benefit $b$.
For $r \gtrsim O(10^{-1})$, the cooperation levels drop as $r$ increases, as shown in Fig.~\ref{fig:finite_mut_c_level}(b).

\begin{figure}
    \centering
    \includegraphics[width=0.9\textwidth]{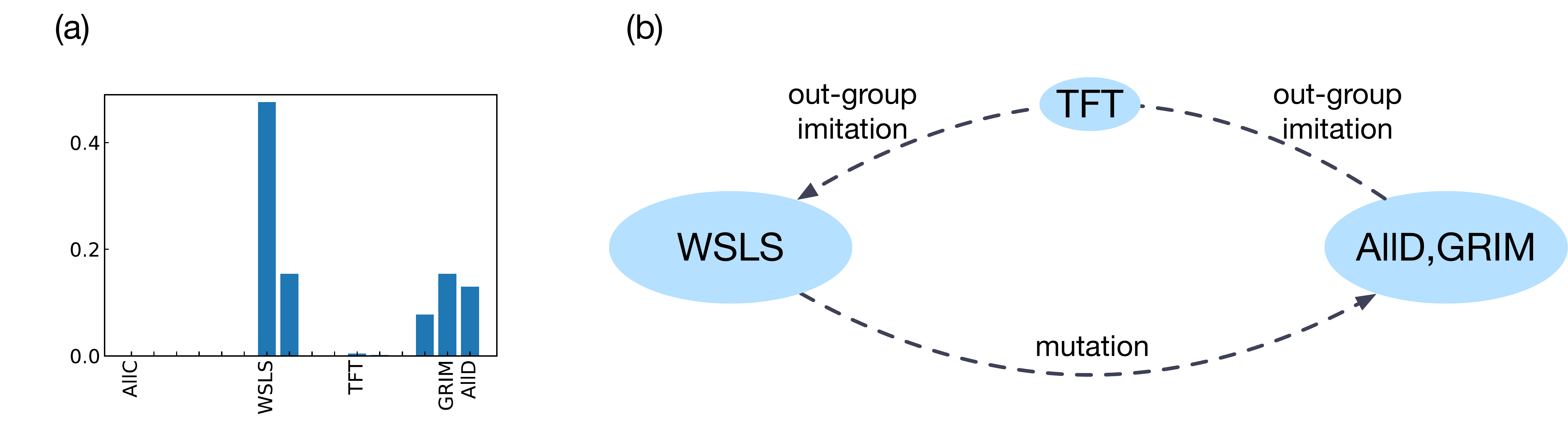}
    \caption{Strategy evolution in group-structured populations when there is a partial separation of time scales. 
    (a) We consider the abundance of each strategy when $b=3$ and $r=0.01$; similar distributions are found for other parameters when $r \lesssim O(10^{-1})$.
    WSLS is most abundant, followed by $S_7$, Grim, AllD, and $S_{13}$. 
    (b) A schematic diagram of the typical transitions between strategies reveals that groups of AllD and Grim players tend to transition towards TFT by out-group imitation. TFT groups in turn tend to transition towards more cooperative strategies like WSLS or AllC.  Finally, WSLS groups tend to transition towards AllD and Grim once these strategies are introduced by mutations.}
    \label{fig:finite_mut_transition}
\end{figure}

To explore these results for group-structured populations in more detail, Fig.~\ref{fig:finite_mut_transition}(a) shows the abundance of strategies in the selection-mutation equilibrium for $b=3$. 
According to this figure, the most abundant strategy is $WSLS$, followed by $S_7$, Grim, AllD, and $S_{13}$ (the latter four strategies are exactly the strategies that have an advantage when directly competing with WSLS, see  right-most column of Table~\ref{tab:memory-1}). 
The underlying evolutionary dynamics are schematically depicted in Fig.~\ref{fig:finite_mut_transition}(b).
Individuals in groups with non-cooperative strategies (such as Grim and ALLD) tend to adopt more cooperative strategies like TFT by out-group imitation. Once such groups contain a TFT-player, TFT may reach fixation by intra-group imitation (TFT is neutral when there is only a single TFT player in the group, and it is selectively favored when there are two TFT players or more). 
TFT-groups in turn are easily replaced by strategies that are more cooperative in the presence of errors, such as AllC and WSLS. 
WSLS groups are comparably stable; as they reach the maximum payoff against themselves, individuals in these groups are unlikely to learn non-cooperative strategies by out-group imitation. However, strategies like AllD and Grim may invade a group of WSLS players once they are introduced by mutations. Assuming that the group is small (the figure depicts the case of $N=2$), AllD and Grim are both likely to take over, thereby closing the evolutionary cycle.
Importantly, the above arguments do not depend on the precise value of $b$; they only depend on the win-lose relationships between strategies. This argument can thus explain why we observe a coexistence between WSLS and non-cooperative strategies for a wide range of benefit values.

The above argument also explains the dependency of cooperation levels on $r$.
When $r$ is sufficiently small, the abundance of TFT also becomes as small as $O(r)$:
The transitions between WSLS and the defectors thus balance, and their abundance as well as cooperation levels do not show significant change as $r$ varies.
However, because the flow from WSLS to the defectors are mainly driven by the mutation events, the abundance of WSLS players drops as mutation events get more frequent.

When $r$ or $M$ is even smaller, evolutionary results become closer to the results observed when there is a complete separation of time scales. 
The crossover point depends on the strength of selection.
When selection strengths are sufficiently weak, the evolutionary dynamics get closer to neutral selection, where fixation times are relatively short.
Here, either WSLS or AllD may happen to take over the entire population with a non-negligible frequency, and the evolutionary dynamics are better described by the complete separation of time scale.

\section*{Summary and Discussion}

Herein, we propose and study a model for the evolutionary dynamics of direct reciprocity in group-structured populations.  
In our model, individuals can adopt new strategies in three different ways. 
They may imitate members of their own group (with whom they also engage in prisoner's dilemma interactions); 
they may imitate members from a different group (with whom they do not interact directly); 
or they may adopt a new strategy by random strategy exploration (similar to a mutation). 
While we derive the model in general terms, we focus on two special cases in particular. 
In the first case, there is.a complete separation of time scales. 
Here in-group imitation occurs much more often than out-group imitation, which in turn occurs more often than mutations. 
For this case, we can analytically derive the fixation probability, the conditional and unconditional fixation times~(Appendix), and the condition for a strategy to be favored in a pairwise competition. 
In the second case, there is only a partial separation of time scales. 
In-group imitation still occurs most often, but now out-group imitation and mutations happen at a similar rate. 
We explore this case by Monte Carlo simulations, and by analyzing the properties of a differential equations that is valid in the limit as the number of groups becomes large. 

For both cases, we explore the effect of group structure by comparing the abundance of cooperation for group-structured populations with the corresponding results for well-mixed populations of equal size. 
Interestingly, we find that the effect of group structure depends on the benefit of cooperation. 
When this benefit is small, group-structured populations are more cooperative than well-mixed populations. 
This ordering reverses once the benefit of cooperation is intermediate or large. 
This result differs from previous models on the evolution of reciprocity in the presence of population structure. 
In van Veelen et al\cite{van2012direct}, the authors explore the evolution of strategies in repeated games with relatedness. 
In their model, there is an assortment parameter $\alpha$ that determines how likely players who use the same strategy are to interact with each other. 
Their simulation results displayed in their Figure~2 suggest that the effect of this kind of population structure is always positive.  
As the assortment parameter $\alpha$ increase, people tend to adopt more cooperative strategies on average. 
In contrast, we find that additional group structure can sometimes prevent the evolution of cooperation. 
This effect occurs because the effect of group structure is ambivalent in our model. 
On the one hand, splitting a well-mixed population into smaller groups promotes cooperation, because people in cooperative groups are more likely 
to act as role models for between-group comparisons. 
On the other hand, group-structured populations lead to smaller effective population sizes, which in turn select for spite\cite{nowak:Nature:2004} and defection\cite{hilbe2018partners}. The overall outcome of these two opposing effects depends on how profitable cooperation is. 

To derive our results, we have focused on a comparably simple strategy space, the space of memory-1 strategies\cite{sigmund:book:2010}. 
One of the open questions is thus the study of strategies with longer memory. 
According to Eq.~(\ref{Eq:pfavored}), there are two requirements for a strategy $\mathbf{p}$ to be successful in group-structured populations.
They need to be efficient ($\pi_{\mathbf{p},\mathbf{p}} \geq \pi_{\mathbf{q},\mathbf{q}}$), but at the same time they should also have a higher payoff than the co-player in a direct interaction ($\pi_{\mathbf{p},\mathbf{q}} \geq \pi_{\mathbf{q},\mathbf{p}}$).
Recent research suggests that strategies exist that satisfy both of these conditions simultaneously. 
These so-called friendly rivals~\cite{do2017combination,murase2018seven,murase2020five,murase2020automata,murase2021friendly} are fully cooperative when they interact with one another, yielding the maximum payoff of $\pi_{\mathbf{p},\mathbf{p}} = b-1$. 
However, they also make sure they are never outperformed by any opponent, $\pi_{\mathbf{p},\mathbf{q}} \geq \pi_{\mathbf{q},\mathbf{p}}$ for all co-player's strategies $\mathbf{q}$. 
It is straightforward to show that when the resident applies a friendly rival strategy $\mathbf{p}$, the fixation probability $\Psi_{\mathbf{q},\mathbf{p}}$ of any mutant is at most $1/MN$. In other words, friendly rivals are evolutionarily robust~\cite{stewart2013extortion} for any environmental condition $N$, $M$, and $b$.
Instantiating a friendly rival strategy, however, requires more than one-round memory~\cite{do2017combination,murase2018seven,murase2020five,murase2020automata,murase2021friendly}. Exploring whether these strategies are particularly favored in group-structured populations is thus an interesting and promising research area for future studies.

Another possible direction is to compare the model with human behavior in empirical or experimental studies.
Group-structured populations could better describe human behavior than the more traditional well-mixed population model because human relationships often involve a limited number of people. 
At the same time, it seems comparably easy in our modern societies to get payoff information even from people that one has no personal ties with. 
It has long been recognized in sociology~\cite{granovetter1973strength}, that such weak links between communities may play a pivotal role, which may also affect the evolution of direct reciprocity.

\section*{Methods}

\subsection*{Monte Carlo simulations}

We use Monte Carlo (MC) simulations to obtain the results in Figs.~\ref{fig:coop_M_dep_N_fixed} and~\ref{fig:low_mut_abundance}.
At each time step, a mutant strategy is randomly selected from the set of deterministic memory-1 strategies (Table~\ref{tab:memory-1}). The resident strategy is replaced by the mutant with the fixation probability calculated by Eq.~(\ref{eq:psi}).
Throughout this paper, we use an error rate $e = 10^{-3}$ and selection strengths $\sigma_{\rm in} = \sigma_{\rm out} = 10$ as our baseline values. 
We conducted the simulations for $10^6$ MC steps while discarding the initial $10^5$ steps, and took the average over five independent runs to obtain a stationary distribution over the strategy space.
While it is possible to obtain the exact stationary distribution of the Markov chains, this method is more vulnerable to underflow.
Instead, we found MC simulations to be more reliable, and our simulations are long enough to ignore the statistical fluctuations.
For numerical calculation of the fixation probability and the fixation times, we use algorithms based on a previous work~\cite{hindersin2019computation}.

For the MC simulations for the scenario with a partial separation of time scales, we proceed as follows. 
First, we prepare a set of randomly selected strategies of size $M$ as an initial state.
For each time step, a group $i$ is randomly selected out of the $M$ groups. A player in this group then mutates with probability $r$. Otherwise, out-group imitation occurs.
In case of a mutation, the mutant strategy is randomly selected from the memory-1 strategy space, and the group $i$ is taken over by the mutant with probability Eq.~(\ref{eq:intra_fixation_prob}).
In case of an out-group imitation, a role model is randomly selected from the groups other than $i$.
The strategy of group-$i$ is then replaced with probability $f_{\mathbf{p} \to \mathbf{q}}^{\rm out} \rho_{\mathbf{q},\mathbf{p}}$ since a single $\mathbf{q}$ player appears in $i$ with $f_{\mathbf{p} \to \mathbf{q}}^{\rm out}$ and the player succeeds in taking over the group with $\rho_{\mathbf{q},\mathbf{p}}$.
We define $M$ Monte Carlo steps as one MC sweep, and the simulations are conducted for $1$ million sweeps. Again, the initial $10\%$ of data are discarded as the initialization period.
The results are averaged over five independent runs.
To make the typical time scale comparable between different $b$, the selection strengths are set to $\sigma_{\rm in} = \sigma_{\rm out} = 30/(b-1)$.
We used OACIS to manage simulation results~\cite{murase2017open}.

\subsection*{Results on fixation times}

One can also calculate the unconditional and conditional fixation times in a group-structured population~\cite{hauert2014fixation,hindersin2019computation}.
To simplify the analysis, we assume in the following that the intra-group dynamics are fast enough such that each group  can be considered as homogeneous.
In other words, we study the dynamics triggered by out-group imitation events and we take $1/\mu$ (the number of the inter-group imitation events) as the unit of time.
First, we consider the unconditional fixation time for the inter-group dynamics, which is defined as the average time until the absorbing states (either fixation or extinction) started from the state where there is a single group of mutants.
As a preparation, the probability that $\mathbf{p}$ succeeds in taking over the population from the state where the number of $\mathbf{p}$ groups is $i$ is
\begin{equation}
\phi_i = \frac{\sum_{j=0}^{i-1}\eta^j}{\sum_{j=0}^{M-1}\eta^j }.
\end{equation}
Using this, the unconditional fixation time is given as
\begin{equation}\label{eq:unconditional_fixation_time}
t_{uc} = \phi_1 \sum_{k=1}^{M-1} \sum_{l=1}^{k} \frac{1}{Q_l^+} \eta^{k-l} 
    = \begin{cases}
        \frac{M(M-1) \left\{ 1 + \exp\left[ \sigma_{\rm out} \left(\pi_{\mathbf{q},\mathbf{q}} - \pi_{\mathbf{p},\mathbf{p}} \right) \right] \right\} }{(1 - \eta^M)\rho_{\mathbf{p},\mathbf{q}}} \sum_{l=1}^{M-1} \frac{1- \eta^{l}}{l(M-l)}  & \text{when $\eta \neq 1$} \\
        \frac{(M-1)\left\{ 1 + \exp\left[ \sigma_{\rm out} \left(\pi_{\mathbf{q},\mathbf{q}} - \pi_{\mathbf{p},\mathbf{p}} \right) \right] \right\}}{\rho_{\mathbf{p},\mathbf{q}}} \sum_{l=1}^{M-1}\frac{1}{l} & \text{when $\eta = 1$}
    \end{cases}.
\end{equation}
The conditional fixation time, which is defined as the expected time until the fixation of $\mathbf{p}$ starting from a single $\mathbf{p}$-group conditioned that $\mathbf{p}$ succeeds in taking over the population, is
\begin{equation}\label{eq:conditional_fixation_time}
t_{c} = \sum_{k=1}^{M-1} \sum_{l=1}^{k} \frac{\phi_l}{Q_l^+} \eta^{k-l}
    = \begin{cases}
          \frac{M(M-1)\left\{ 1 + \exp\left[ \sigma_{\rm out} \left(\pi_{\mathbf{q},\mathbf{q}} - \pi_{\mathbf{p},\mathbf{p}} \right) \right] \right\} }{\rho_{\mathbf{p},\mathbf{q}}(1-\eta^M)(1-\eta)} \sum_{l=1}^{M-1} \frac{(1-2\eta^l+\eta^M)}{l(M-l)}   & \text{when $\eta \ne 1$} \\
          \frac{(M-1)^2 \left\{ 1 + \exp\left[ \sigma_{\rm out} \left(\pi_{\mathbf{q},\mathbf{q}} - \pi_{\mathbf{p},\mathbf{p}} \right) \right] \right\} }{\rho_{\mathbf{p},\mathbf{q}}}   & \text{when $\eta = 1$}
    \end{cases}.
\end{equation}

This fixation time can be exceedingly long.
For instance, consider the dynamics between AllC and AllD.
The intra-group dynamics favor AllD while the inter-group dynamics favor AllC.
When there are AllC groups and AllD groups, members in AllD groups tend to consider AllC groups when they engage in out-group imitation. 
However, the AllC player who newly appears in the AllD group immediately disappears again due to the intra-group dynamics.
Thus, AllC fails to spread in AllD groups again and again.
In other words, both $Q_i^{+}$ and $Q_i^{-}$ can be quite small even if $\eta$ ($\equiv Q_i^{-}/Q_i^{+})$ remains finite.
As a result, the population configuration remains the same for a long time, and the fixation times increase dramatically (unless the selection strengths are sufficiently weak).

\bibliography{main}

\section*{Acknowledgements}
Y.M. acknowledges support from Japan Society for the Promotion of Science (JSPS) (JSPS KAKENHI; Grant no. 21K03362, Grant no. 21KK0247, Grant no. 22H00815).
S.K.B. acknowledges support by Basic Science Research Program through the National Research Foundation of Korea (NRF) funded by the Ministry of Education (NRF-2020R1I1A2071670).
Y.M. and S.K.B. appreciate the APCTP for its hospitality during the completion of this work.
C.H. acknowledges generous funding from the European Research Council (ERC) under the European Union's Horizon 2020 research and innovation program (Starting Grant 850529: E-DIRECT).

\section*{Author contributions statement}

Y.M. and C.H. designed the research, Y.M. and S.K.B. conducted analysis, and Y.M. carried out the simulation. All authors wrote and reviewed the manuscript.

\section*{Additional information}

The authors declare no competing interests.

\section*{Data Availability}
The source code for this study is available at \url{https://github.com/yohm/sim_grouped_direct_reciprocity}.

\end{document}